\documentstyle[11pt,newpasp,epsf,twoside]{article}
\def\edcomment#1{\iffalse\marginpar{\raggedright\sl#1\/}\else\relax\fi}
\reversemarginpar

\begin{document}
\title{Chemical Abundances and the Hierarchical Clustering}
 \author{Tissera, P. B.}
\affil{Institute of Astronomy and Space Science, Conicet}
\author{Lambas, D. G.}
\affil{IATE, Observatorio Astron\'omico C\'ordoba and Conicet}

\begin{abstract}
We studied the chemical enrichment of the interstellar medium and
stellar population of the building blocks of current typical galaxies in 
the field, in cosmological hydrodynamics simulations. The simulations include
detail modeling of chemical enrichment by SNIa and SNII. 
In our simulations the metal missing problem is caused by chemical elements 
being
locked upon in the central regions (or bulges) mainly, in stars.
 Supernova energy feedback could help to reduce this
concentration by expelling metals to the intergalactic medium.
\end{abstract}

We studied the chemical enrichment history of the structure in 
a hierarchical scenario by using a cosmological SPH code which
includes detail chemical enrichment by SNIa and SNII (Mosconi et al. 2001).
In order to compare with available observational data we draw random
line-of-sight (LOS) through the building blocks of 
current typical galaxies in the field. We estimated the HI densities 
columns and chemical abundances of different elements as a function 
of redshift. We studied the column densities that satisfied the observational
criteria used for Damped Lyman Systems. 
Our results show that when mapped by random LOS the gaseous components
in building blocks
show the same level of metallicity enrichment ($\approx -0.30 $) and
$\alpha$-enhancement as those estimated for the
observed DLAs (Tissera et al. 2002; Cora et al. 2003) .
The star formation rate per unit are in the simulated DLAs are, on 
averaged, an order of magnitude lower than those of the whole system
(Tissera \& Lambas 2004). 
Comparing the mean abundances of the gaseous components
and the stellar populations along
the LOS and those of the whole galactic systems, we found that the
simulated DLAs unpredicted the abundances by $\approx 40-  70\%$.
The missing metals are located in the central regions of the simulated 
galactic systems and mostly locked up in stars. 
These regions are mapped with a lower probability
by LOS because of geometrical effects.
Our results suggest that the metal missing problem (Pettini 2003)
could be produced by chemical elements being locked stars and in the central
regions. However, we note that 
 SN galactic wind could be able  expelled a significant fraction to the
intergalactic medium reducing the problem.

\begin{figure}
\plotfiddle{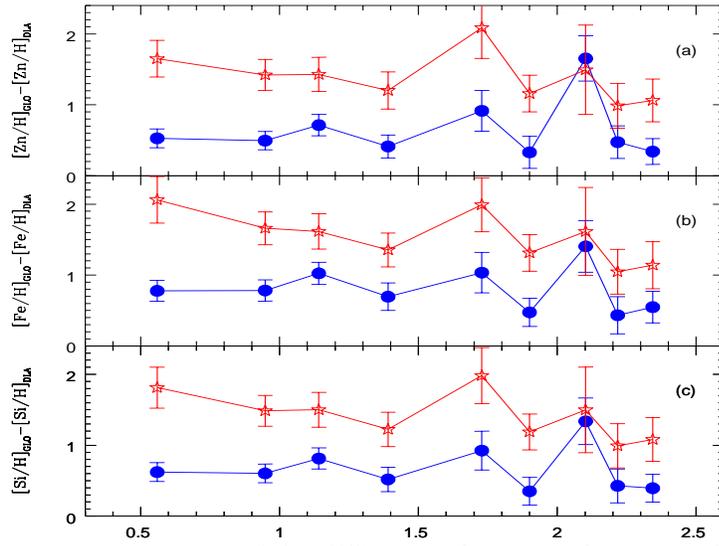}{2.4in}{0}{50}{40}{-160}{-85}
\caption{Averages of the  differences between the mean abundances of
the gaseous components (filled circles) and the stellar populations
(open stars) in 
the simulated DLAs and those of their hostes systems as a function
of redshift. Bootstrap errors shown.
}
\end{figure}


\end{document}